\documentclass[twocol]{ametsocV6.1}

\usepackage{comment}

\title{Online Test of a Neural Network Deep Convection Parameterization in ARP-GEM1}

\authors{Blanka Balogh,\aff{a}\correspondingauthor{Blanka Balogh, blanka.balogh@meteo.fr} 
David Saint-Martin,\aff{a} 
Olivier Geoffroy,\aff{a}}

\affiliation{\aff{a}{CNRM, Université de Toulouse, Météo-France, CNRS, Toulouse, France}}

\abstract{In this study, we present the integration of a neural network-based parameterization into the global atmospheric model ARP-GEM1, leveraging the Python interface of the OASIS coupler. This approach facilitates the exchange of fields between the Fortran-based ARP-GEM1 model and a Python component responsible for neural network inference. As a proof-of-concept experiment, we trained a neural network to emulate the deep convection parameterization of ARP-GEM1. Using the flexible Fortran/Python interface, we have successfully replaced ARP-GEM1's deep convection scheme with a neural network emulator. To assess the performance of the neural network deep convection scheme, we have run a 5-years ARP-GEM1 simulation using the neural network emulator. The evaluation of averaged fields showed good agreement with output from an ARP-GEM1 simulation using the physics-based deep convection scheme. The Python component was deployed on a separate partition from the general circulation model, using GPUs to increase inference speed of the neural network.} 

\begin{document}

\maketitle

\statement 
In this study, we used the Python interface of the OASIS coupler to realize field exchanges between a Fortran-based General Circulation Model (GCM) and a Python-based component responsible for the inference of a neural network (NN) parameterization. By leveraging a heterogeneous architecture — running the GCM on CPU nodes and the NN component on GPU nodes — we achieved significant improvements in execution speed. The resulting GCM, with the NN replacing the physical deep convection parameterization, demonstrated stability for at least five years, highlighting the viability of integrating NN-based components into GCMs.

\section{Introduction} 
The application of Artificial Intelligence (AI), especially the use of neural networks (NNs) in Earth System Modeling has expanded significantly over the past decade, particularly in the realm of Numerical Weather Prediction (NWP). Since 2022, following the pioneering work of \cite{keisler_forecasting_2022}, which introduced a Graph NN designed to emulate ERA-5 \citep{hersbach_era5_2020} at a horizontal resolution of 1.5 degrees, there has been increasing interest in using AI to emulate entire atmospheric models, even at higher resolutions \citep[e.g.,][]{pathak_fourcastnet_2022, bi_pangu-weather_2022, lam_graphcast_2022, lang_aifs_2024}.

While these approaches hold significant promise for NWP, their application to climate modeling remains challenging. Firstly, climate models require rigorous guarantees of numerical stability. In NWP, predictions typically span from 5 to 15 days, but climate models are designed to simulate weather patterns over several decades to millennia. This extended time horizon necessitates extreme robustness, as even small numerical instabilities can compound and render long-term simulations inaccurate. Secondly, there is a lack of a suitable `ground truth' dataset in climate modeling, akin to the ERA-5 reanalysis in NWP. ERA-5 integrates observational data into the IFS simulations via data assimilation, making it one of the most reliable datasets for global-scale, spatialized observations. In contrast, the latest CMIP (Coupled Model Intercomparison Project) exercise, particularly CMIP-6 \citep{eyring_overview_2016}, offers data that represents a spread across different models, but is not grounded in observational truth. Currently, the CMIP ensemble remains the closest equivalent of the ERA-5 dataset in climate modeling (i.e., including data to describe future climates). However, training AI systems on CMIP data may train the models underlying biases present in the original climate models. Lastly, interpretability remains a critical concern. In climate science, understanding the underlying causes of modeled phenomena is essential, but fully AI-based atmospheric models can obscure this insight. Although significant progress has been made in improving AI transparency \citep[e.g.,][]{mcgovern_making_2019, toms_physically_2020}, extracting meaningful physical knowledge from NN-based models remains a complex task, particularly when considering the intricate interactions that drive Earth's climate system. 

Recent advances in AI-driven Earth system modeling, particularly whole-model emulation, have tended to overshadow progress in hybrid modeling, especially the development of data-driven subgrid-scale parameterizations. These data-driven approaches aim to enhance specific components of climate models, either by improving the accuracy of existing physical parameterizations or by accelerating numerically expensive schemes. Subgrid-scale parameterizations, which represent processes occurring at smaller scales than the model grid, often target areas such as convection \citep[e.g.,][]{rasp_deep_2018, gentine_could_2018, brenowitz_prognostic_2018, brenowitz_spatially_2019, ogorman_using_2018, beucler_towards_2020, heuer_interpretable_2024}, gravity wave drag \citep[e.g.,][]{chantry_machine_2021}, radiative transfer \citep[e.g.,][]{krasnopolsky_using_2013, ukkonen_exploring_2022}, or even full subgrid-scale physics \citep[e.g.,][]{hu_stable_2024}. Other approaches focus on bias correction, aiming to rectify systematic errors within the models \cite[e.g.][]{yuval_stable_2020, yuval_use_2021}. 

One of the key challenges inhibiting further advancement in this area is the complexity of interfacing General Circulation Models (GCMs) with NNs. Most current GCMs are written in Fortran and are optimized to run on hundreds of Central Processing Units (CPUs). In contrast, NNs are typically developed using Python-based frameworks, such as PyTorch \citep{paszke_pytorch_2019} and TensorFlow \citep{abadi_tensorflow_2015}, which are designed for efficient execution on Graphics Processing Units (GPUs). This technological gap has made it difficult to integrate NNs directly into GCM workflows, contributing to the scarcity of online NN testing in GCMs, often in a simplified, aquaplanet configuration \citep[e.g.][]{gentine_could_2018, yuval_stable_2020}.

In this paper, we introduce a coupling method for the global atmosphere model ARP-GEM1 \citep{saint-martin_arp-gem_2024, geoffroy_arp-gem_2024}, a computationally efficient version of the ARPEGE/IFS atmospheric model.
Our interface enables online testing of any data-driven parameterization written in Python, thus bridging the technology gap between GCMs and modern AI frameworks. In Section \ref{sec:interfacing}, we detail the choice of a Fortran/Python interface to use NN-based parameterizations in ARP-GEM1. Section \ref{sec:NN} describes a realistic test case, where we train a NN to emulate ARP-GEM1's existing deep convection parameterization. Online performance of the NN is evaluated in the last section, using data from an ARP-GEM1 simulation where the NN replaced the physical parameterization of deep convection.

\section{Fortran/Python interfacing approaches}
\label{sec:interfacing}

\begin{figure*}
\centerline{\includegraphics[width=28pc]{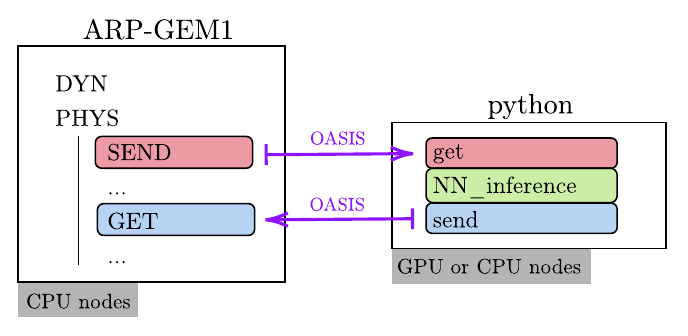}}
    \caption{Field exchange between ARP-GEM1 and the Python-based NN inference component. In the online experiment, ARP-GEM1 computes the dynamics tendencies first. During the physical tendencies computation (including subgrid-scale parameterizations), ARP-GEM1 sends the necessary NN input fields to the Python component. The Python process performs the NN inference and sends back the NN output tendencies for deep convection. These tendencies are incorporated into the total tendencies in ARP-GEM1, which are then integrated and used by the dynamics at the next timestep.}
    \label{fig:arp_oasis_xios}
\end{figure*}

The Fortran programming language has been widely used to develop numerical models of the atmosphere due to its computational efficiency on high-performance computing (HPC) systems. Consequently, most GCMs are based on millions of lines of legacy Fortran code. While interfaces between Fortran and C/C++ are well established, integrating Fortran with Python, the language in which most machine learning (ML) developments are conducted, remains a challenge. Hence, the lack of an efficient interface between Python and Fortran is the primary limitation in using data-driven parameterizations in GCMs. Although NN inference is optimized in Python, particularly for GPU-based execution, using GPUs to execute Fortran-based numerical models often require significant changes to the models' architecture. This complexity adds further difficulty to the efficient integration of ML-based parameterizations. To address the interface gap between Fortran and Python, several methods have been developed in recent years, broadly categorized into two approaches: bindings and MPI (Message Passing Interface)-based field-exchange methods.

Bindings allow NN inference to be called directly within Fortran code via the Fortran/C interface. The use of this packages requires to save the NN model into specific numerical formats. Notable examples of such packages include Forpy\footnote{\url{https://github.com/ylikx/forpy}}, the Fortran-Keras Bridge \cite{ott_fortran_2020}, ECMWF’s Infero\footnote{\url{https://github.com/ecmwf/infero}}, and FTorch\footnote{\url{https://github.com/Cambridge-ICCS/FTorch}}. These methods are advantageous due to their simplicity and computational efficiency. Often, only a few lines of code need to be added to the existing numerical models, allowing for seamless NN inference execution. In this setup, the NN is run on the same computational resources as the Fortran-based GCM, typically CPU nodes on HPC systems. However, these binding methods are often tied to specific Python packages, limiting their flexibility. Additionally, bindings tend to support only basic operations, such as NN inference, which may restrict their broader application.

The second category of interfacing methods relies on coupling the Fortran-based GCM with a Python-based component, using MPI to exchange fields between the GCM and the Python component, including the NN inference. Examples of interface packages based on field exchange include OASIS developed by CERFACS \citep{craig_development_2017}, PhyDLL-0.2 also by CERFACS \citep{serhani_graph_2024}, and SmartSim by Cray \citep{SmartSim_2022}. \cite{wang_stable_2022} has developed a custom coupler, called \textit{the NN-GCM coupler}, for the SPCAM using MPI to exchange fields between the GCM and a NN-based parameterization. The key advantage of this method is its flexibility in terms of computational resources, as any operation written in Python can be executed within the component. Notably, this setup does not require recompiling the GCM when there are changes to the NN model or its components (e.g., normalization or NN architecture adjustments). The second advantage is that these coupling libraries can already be present in atmospheric models, enabling the exchange and interpolation of fields between different components (e.g., atmosphere, ocean, surface) of a coupled system, thereby facilitating code maintenance. For example, the OASIS coupler is already used in several climate models \citep[e.g.,][]{williams_met_2018, mauritsen_developments_2019, voldoire_cnrm-cm6-1_2019, boucher_presentation_2020}. Moreover, the Fortran-based GCM and the Python component can run on separate partitions of the HPC system. For instance, the GCM can be run on CPU nodes while the NN inference can be executed on GPUs. However, this approach also has notable drawbacks. Field exchanges via MPI can be computationally expensive. For optimal numerical efficiency, the Python component should run in parallel with other processes within the GCM. As a result, successful implementation of this interface demands a deep understanding of the GCM's structure. Furthermore, many current GCMs are already coupled with other components, such as ocean or land models, via MPI.

We chose to implement a method from the field-exchange family, specifically the Python interface of the OASIS coupler, version OASIS3-MCT\_5.0\footnote{\url{https://gitlab.com/cerfacs/oasis3-mct/-/tree/OASIS3-MCT_5.0}}, and its Python interface, pyOASIS \citep{pyoasis_doc}.
The interfacing with OASIS follows the strategy described in \cite{voldoire_surfex_2017}. Initialization and finalization steps are similar, while the multi-process partition definition and listing of exchanged fields have been slightly adapted. The receiving and sending of coupling fields are described in Figure \ref{fig:arp_oasis_xios}. ARP-GEM1 transmits the necessary physical fields for NN inference, before the Python component sends the output from the NN to ARP-GEM1. The field exchange between ARP-GEM1 and the Python component is performed during the computation of the 'physics' in ARP-GEM1. The choice of the method and the OASIS coupler was motivated by several key factors. First, the flexibility offered by running operations in a Python component once the coupling is implemented is particularly beneficial for new researchers who may not be familiar with Fortran. Secondly, since ARP-GEM1 is designed to run on CPU nodes only, using the field-exchange approach allows us to offload the Python component to GPUs, leveraging the speedup potential for tensor-based operations such as normalization and NN inference. Note that our implementation also enables to run the Python part on CPUs. Lastly, the OASIS coupler is already integrated into ARP-GEM1 and allows to couple the atmospheric model with other components (e.g., ocean model), making it a natural choice for our application.

\section{Example of a NN parameterization}
\label{sec:NN}

\begin{table}[h]
\centering
\begin{tabular}{l c}
\hline
 \textbf{Input variable}  & \textbf{Dimension}  \\
  \cline{1-2}
  $T$  & 50   \\
  $q_T$  & 50   \\
  $w$  & 50   \\
  $P_s$  & 1   \\ 
  $LSM$  & 1   \\
  $LHF$  & 1   \\
  $SHF$  & 1  \\
  \hline
  \textbf{TOTAL} & \textbf{154} \\
\hline
\end{tabular}
\quad
\hspace*{.5cm}
\begin{tabular}{l c}
\hline
 \textbf{Output variable}  & \textbf{Dimension}  \\
\hline
  $\partial_t s^{(\mathrm{NN})}$  & 44   \\
  $\partial_t q^{(\mathrm{NN})}$  & 44   \\
  \hline
  \textbf{TOTAL} & \textbf{88} \\
\hline
%\vspace*{1.55cm}
\end{tabular}
\vspace*{.5cm} %maybe not needed
\caption{Inputs and ouputs of the NN emulator of the deep convection parameterization.}
\label{tab:inp_out}
\end{table}

The main objective is to perform a technical test of the implementation of the Fortran/Python interface using OASIS. Furthermore, this study also serves as a preliminary test of replacing physical parametrizations by NNs within this idealized framework.

\subsection{Training the NN}
The deep convection scheme of ARP-GEM1 is based on \cite{tiedtke_representation_1993}, revised by \cite{bechtold_advances_2008, bechtold_representing_2014}. A full description is given in \cite{ecmwf-2019}, with additional modifications in \cite{saint-martin_arp-gem_2024}. The scheme will be referred to as Tiedtke-Bechtold. The parameterization outputs tendencies of moist static energy $\partial_ts$, specific humidity $\partial_tq$ and wind components. For the sake of simplicity, the NN only learns $\partial_t s$ and $\partial_t q$. The momentum tendencies for deep convection remains computed with the Tiedtke-Bechtold scheme.

We use an ARP-GEM1 simulation to generate the dataset used to build the learning sample. In this study, ARP-GEM1 operates at a horizontal resolution of TCo179, using a cubic-octahedral reduced Gaussian grid (approximately 55-km resolution) and 50 hybrid model levels, extending from the surface up to 2 hPa. The model timestep is set to $\Delta t=900$ seconds. For computational efficiency, the radiative transfer scheme is applied on a coarser horizontal grid (approximately 150 km resolution), and only every eight timesteps. ARP-GEM1 models atmospheric processes exclusively, with sea surface temperatures and other external forcings prescribed.

To build the training dataset, we used data from a one-year simulation of ARP-GEM1, with forcings for the year 2005. Model outputs and inputs used to fit the NN were saved every three hours. The input variables of the NN are: vertical profiles of temperature $T$, specific total water content $q_t$ and the large-scale vertical component of wind $w$ as well as the following 2D variables: surface pressure $P_s$, land-sea mask $LSM$, latent heat flux $LHF$, and sensible heat flux $SHF$. The NN emulator outputs vertical profiles dry static energy tendency $\partial_t s^{(\mathrm{NN})}$ and specific humidity tendency $\partial_t q^{(\mathrm{NN})}$ from deep convection processes only. These variables and their dimensions are listed in Table \ref{tab:inp_out}. Since the deep convection scheme does not operate on the top six vertical levels (which represent the stratosphere and the mesosphere), we excluded these levels, limiting the target variables to the remaining 44 vertical levels. Thus, the NN output size is $2 \times 44 = 88$, reflecting the moistening and heating tendencies on these levels. To generate the training dataset, we randomly select 20,000 atmospheric columns out of a total of 136,000 columns, for each saved timestep from the one-year simulation. This results in a total training dataset of approximately 48 million samples (or columns), with 154 input features and 88 output features for the NN. The dataset is then randomly split into training (85\%) and validation (15\%) subsets, with random shuffling applied at the beginning of each epoch to limit overfitting.

We train a fully connected feedforward NN with six layers, each containing 1024 nodes, using PyTorch \citep{paszke_pytorch_2019}. The model contains approximately 6.5 million trainable parameters. Note that we tested several architectures, including recurrent and 1D convolutional layers, as well as models with separate branches for vertical profiles and scalar variables (not shown). All architectures produced similar results. Consequently, we opted for the simplest model for the online testing phase. The activation function used is ReLU, and the loss function is Mean Squared Error (MSE). The NN was trained over 50 epochs using a single V100 GPU, requiring a total of 20 GPU-hours. During training, we monitored and minimized the loss on the validation set to avoid overfitting. Only the model weights and biases that minimized validation loss were saved for further use.

\subsection{Offline evaluation}

\begin{figure}
\centerline{\includegraphics[width=19pc]{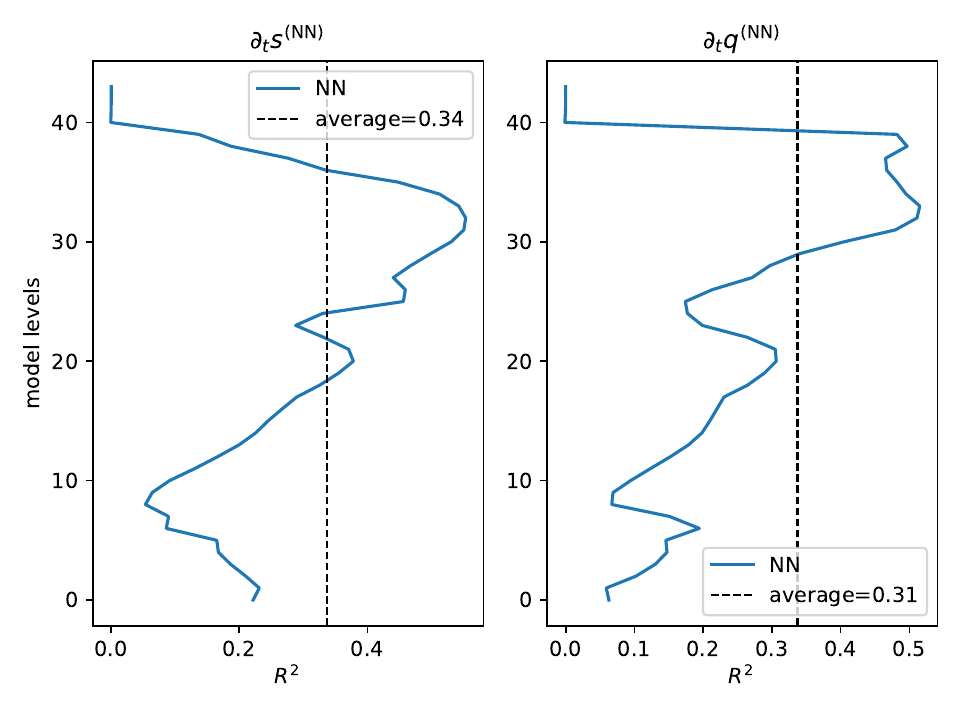}}
    \caption{Offline evaluation of the NN. $R^2$ scores were computed using the test sample. Global $R^2$ score of the model is 0.33.}
    \label{fig:NN_offline}
\end{figure}

\begin{figure}
\centerline{\includegraphics[width=19pc]{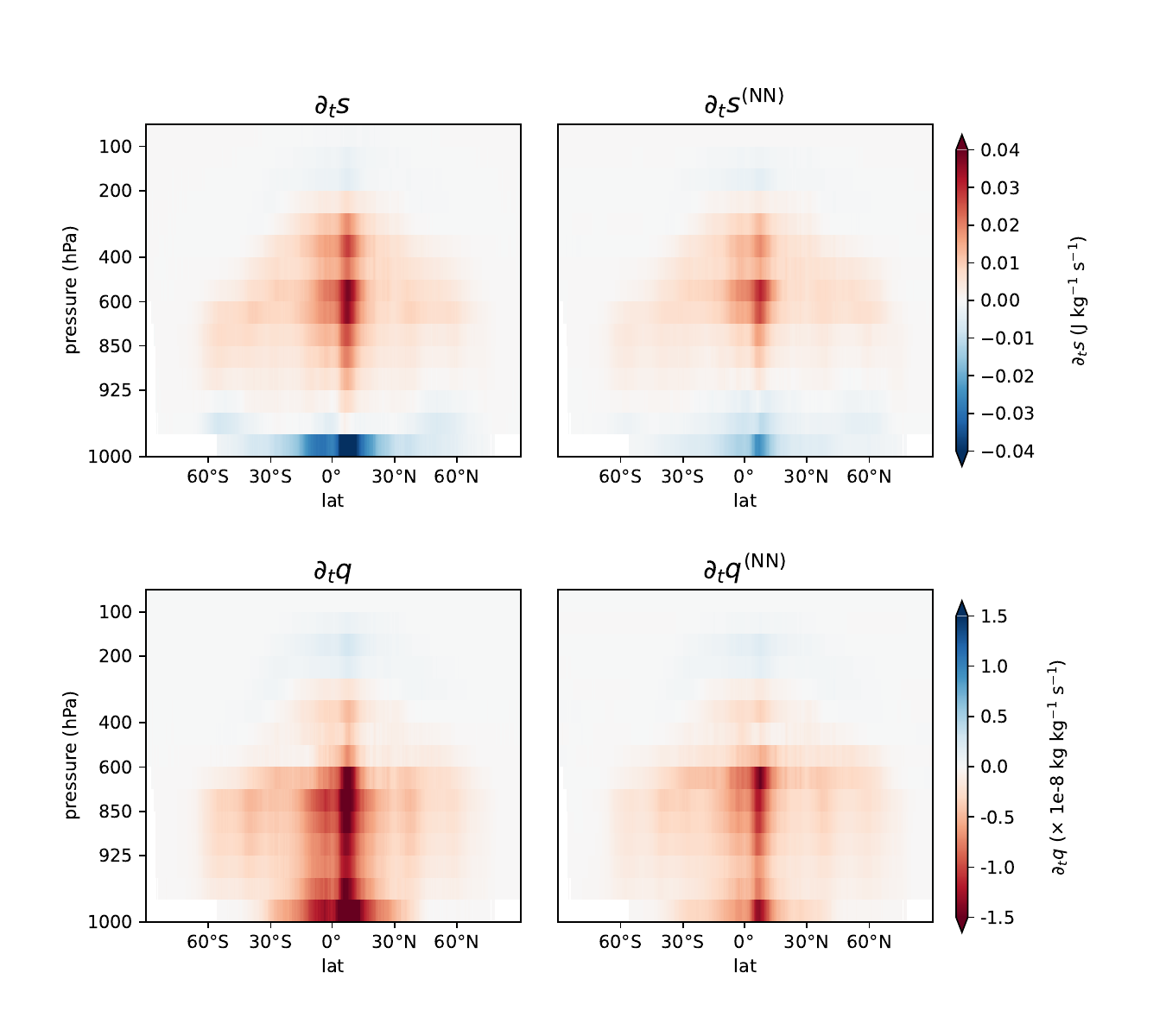}}
    \caption{Zonal average values of tendencies computed on the offline validation sample, for the offline test period (01 Jan 2006 to 31 Dec 2006). Data has been interpolated from model levels to pressure levels. The left column shows the target variables from the validation sample, while the right column presents the NN inference results on the same validation sample. The top row represents dry static energy tendencies, $\partial_t s$, and the bottom row displays specific humidity tendencies, $\partial_t q$, both from deep convection processes.}
    \label{fig:zm_offline}
\end{figure}

The offline validation is done using data from an ARP-GEM1 simulation, with forcings for the year 2006, ensuring that the validation data is independent of the training data (which is based on year 2005). The offline validation sample is built by saving all the columns, every 12h. The NN inference is then applied to the whole dataset. The validation sample size is approximately $N_{test} = 100 \times 10^{6}$ samples (or columns).  

Using the output from the 2006 simulation, we compute the global $R^2$ score for each vertical level $k$, as defined by the following equation:
\begin{equation}
    R^2(k) = \frac{1}{N_{test}} \sum_{i=1}^{N_{test}} \left( 1 - \left( \frac{y_i(k) - y_i^{(\mathrm{NN})}(k)}{y_i(k) - \overline{y}(k)}\right)^2 \right),
\end{equation}
where $y_{1 \leq i \leq N_{test}}(k)$ are the target variables at vertical level $k$, $y_{1 \leq i \leq N_{test}}^{(\mathrm{NN})}(k)$ are the corresponding NN predictions, and $\overline{y}(k)$ is the mean of the target variables at level $k$. The vertical profiles of the $R^2$ scores are presented in Fig. \ref{fig:NN_offline}. The vertical average $R^2$ score for heating tendencies is 0.34 and is 0.31 for moistening tendencies. Note that to maintain consistency with the configuration used for the subsequent online tests, we only considered the NN output tendencies on the 40 lowest vertical levels, masking the four highest levels in the output. Although these $R^2$ values might seem modest, the NN effectively captures key atmospheric structures. This is evidenced in Fig. \ref{fig:zm_offline}, which shows the zonal mean tendencies of dry static energy and specific humidity derived from the 2006 test dataset. The NN emulator successfully replicates the general structure of the tendencies, though with some smoothing effects. Note that $\partial_t s$ and $\partial_t q$ output from the Tiedtke-Bechtold scheme are noisy, which can easily lead to a `double penalty' issue. This can explain the low point-by-point $R^2$ scores associated with the NN output. Thus, a more accurate evaluation of the NN's performance requires an online assessment, which will be the focus of the next section.

\section{Online performance of the NN}
\label{sec:online}

\subsection{Implementation of the NN in ARP-GEM1}

The aim of this section is to describe the results of simulations where the NN described in Section \ref{sec:NN} replaces part of the Tiedtke-Bechtold scheme in ARP-GEM1 (Fig. \ref{fig:arp_oasis_xios}). The NN emulator outputs vertical profiles of dry static energy tendency $\partial_t s^{\mathrm{(NN)}}$ and specific humidity tendency $\partial_t q^{\mathrm{(NN)}}$ from subgrid-scale deep convection processes. The input and output field (Table \ref{tab:inp_out}) exchange occurs at every model timestep, except for the constant field land-sea mask, which is sent only at the first timestep. 

Since the NN does not directly output convective precipitation, it must be computed from the moistening tendencies $\partial_t q^{\mathrm{(NN)}}$ generated by the NN. For a single atmospheric column, subgrid-scale deep convection surface precipitation rate ($\mathcal{P}_{\mathrm{DC}}$) can be approximated as the vertical integral of the convective moistening tendencies, estimated using the following equation:
\begin{equation}
    \mathcal{P}_{\mathrm{DC}}^{\mathrm{(NN)}} = \max(0, \sum_{k \leq N_{lev}+1} \rho_k \Delta z_k \partial_t q_{k}^{\mathrm{(NN)}}),
\end{equation}
where $N_{lev}$ is the number of vertical levels, $\rho_k$ the air density at vertical level $z_k$ and $\Delta z_k=z_{k-1/2}-z_{k+1/2}$ represents the thickness of the $k^{\text{th}}$ atmospheric layer. For physical consistency, negative precipitation values are thresholded to 0. 
Note that such constraints could be directly incorporated into the NN. To distinguish between liquid (rain) and solid (snow) precipitation, we use the atmospheric temperature at the lowest vertical level: if the temperature is below freezing, the precipitation is classified as  snow for that column; otherwise, it is treated as rain from deep convection.

\subsection{Online evaluation}
To evaluate the online performance of the NN deep convection parameterization, we conducted three experiments, all of which spanned the same five-year period from January 1, 2006, to December 31, 2010. 
\begin{itemize}
    \item \textbf{ARP-GEM1}: a reference run, with ARP-GEM1's physical parameterization of deep convection (Tiedtke-Bechtold scheme). 
    \item \textbf{ARP-GEM1-NNDC}: ARP-GEM1 simulation where the deep convection parameterization for $\partial_t s$ and $\partial_t q$ has been replaced with the NN parameterization described in Section \ref{sec:NN}. The momentum tendencies are still computed by the Tiedtke-Bechtold scheme.
    \item \textbf{ARP-GEM1-noDC}: ARP-GEM1 simulation without deep convection parameterization.
 This experiment allows to highlitht the role of the Tiedtke-Bechtold scheme in the model. It serves as a additional benchmark to evaluate the performance of the NN-based emulator. Note that only humidity and temperature tendencies are set to zero in this experiment; the momentum tendencies are remains computed by the Tiedtke-Bechtold scheme.       
\end{itemize}

\begin{figure*}
\centerline{\includegraphics[width=28pc]{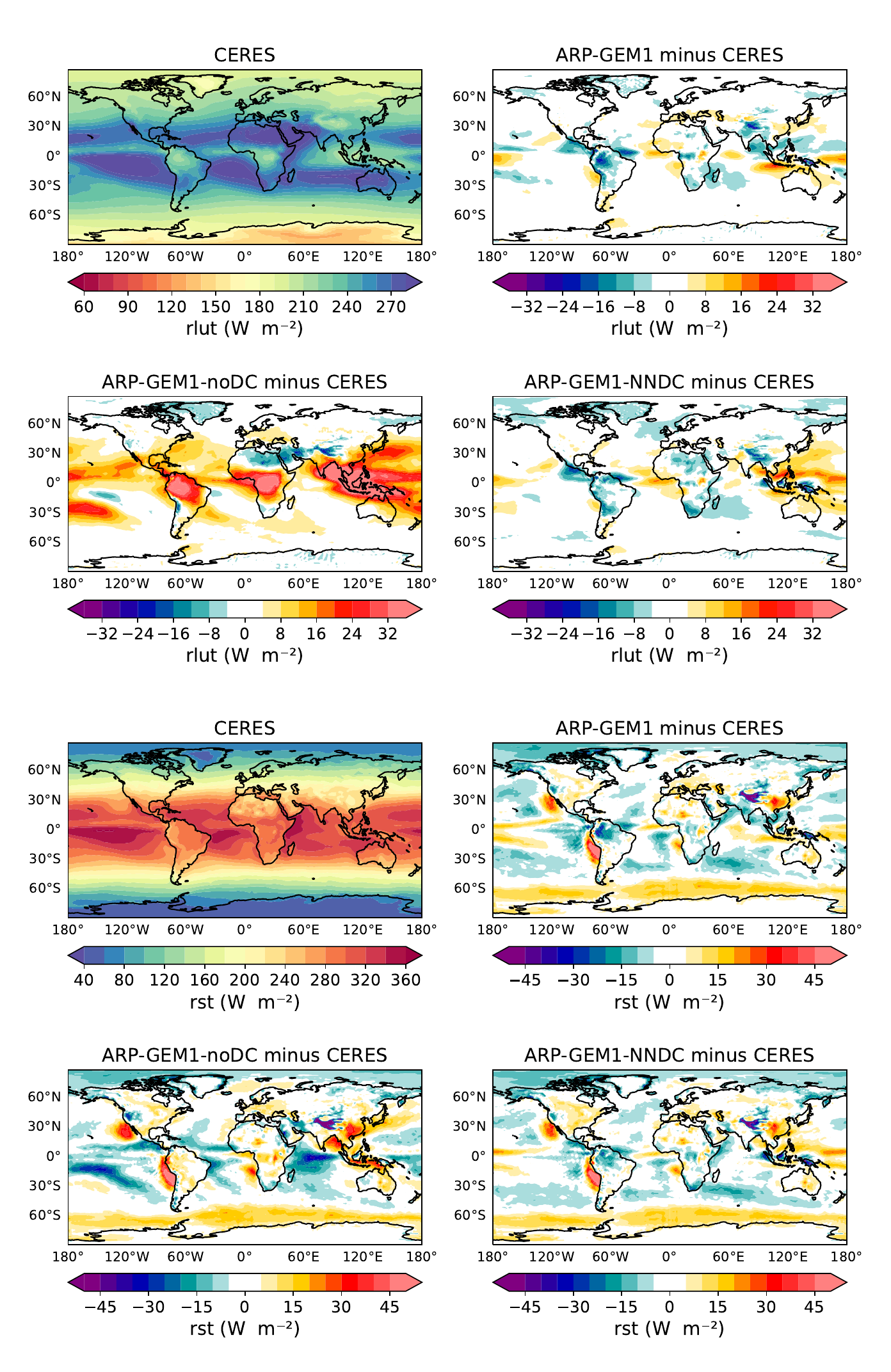}}
    \caption{CERES observations regridded to TCo179 resolution and anomalies of outgoing longwave radiation (rlut, top half) and net shortwave radiation (rst, bottom half). The average values are computed from simulations covering a 5-year period from 01 Jan 2006 to 31 Dec 2010, from different ARP-GEM1 simulations (see Section \ref{sec:online}).} 
    \label{fig:rlut}
\end{figure*}

\begin{figure*}
\centerline{\includegraphics[width=28pc]{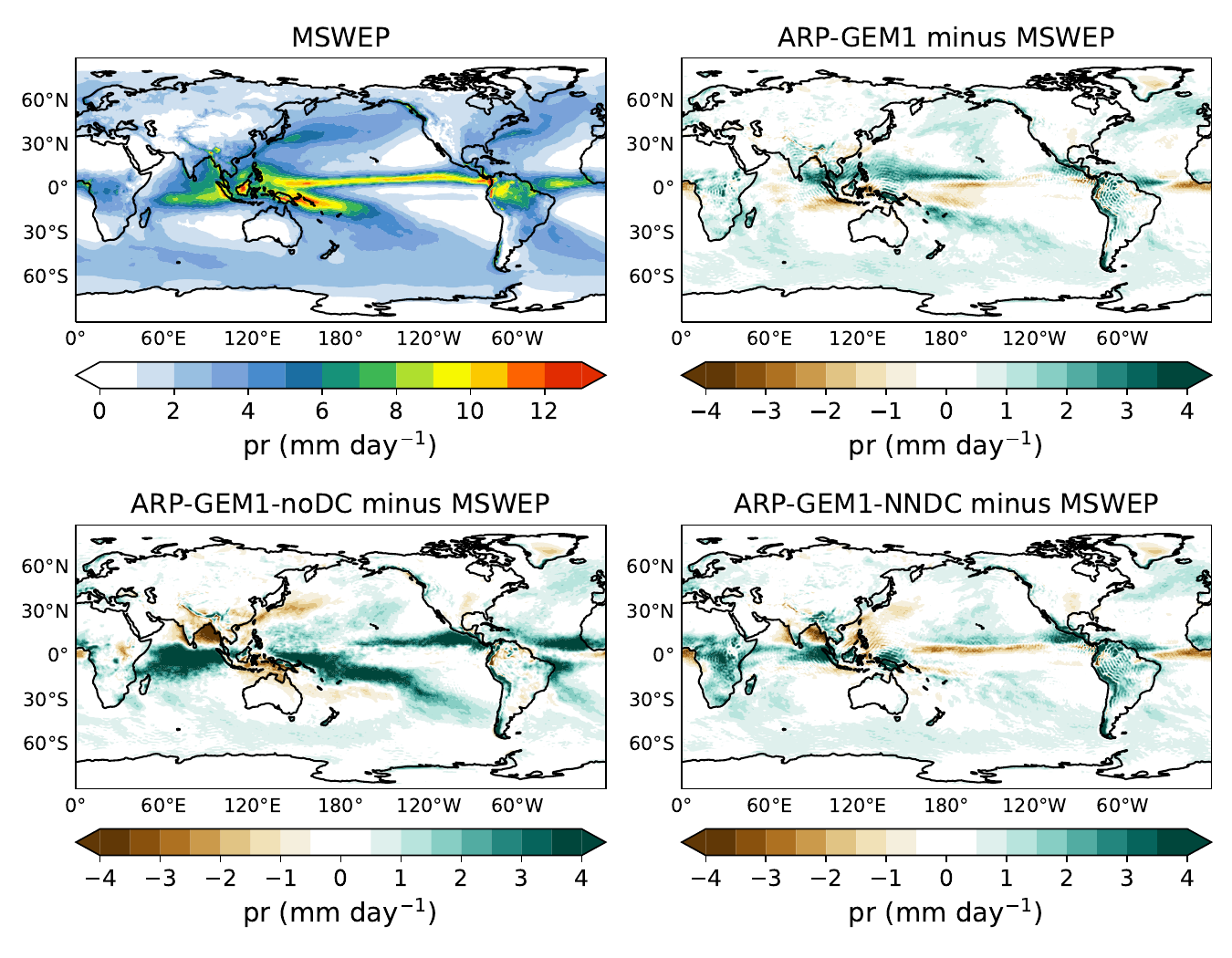}}
    \caption{MSWEP observations regridded to TCo179 resolution and total precipitation anomalies. The average values are computed from simulations covering a 5-year period from 01 Jan 2006 to 31 Dec 2010, from different ARP-GEM1 simulations (see Section \ref{sec:online}).}
    \label{fig:precip}
\end{figure*}

For model evaluation, we focus on top of the atmosphere (TOA) radiation and precipitation fields. Energy budget variables can be considered as main variables to evaluate any deviation from of the model energetical equilibrium associated with the NN. Figure \ref{fig:rlut} provides a comparison of CERES dataset \citep{loeb-2009}, regridded to match ARP-GEM1's horizontal grid, with the output from the ARP-GEM1 experiments. The figure displays outgoing longwave (LW) radiation flux (OLR) and net shortwave radiation. The precipitation field is directly related to deep convection. It is also a strong indicator of the model equilibrium, view that convection play a key role in shaping large-scale circulation. We focus here on total precipitation, that include as well  large-scale precipitation and shallow convection precipitation. A comparison of precipitation with the Multi-Source Weighted-Ensemble Precipitation (MSWEP) dataset \citep{beck_mswep_2017} is given by Figure \ref{fig:precip}.

When using the NN emulator to replace the traditional deep convection parameterization in ARP-GEM1, the resulting anomaly patterns for both radiative variables and precipitation closely resemble those of the reference simulation. The primary discrepancies between the ARP-GEM1 and ARP-GEM1-NNDC simulations are observed within the Pacific Warm Pool. This region, characterized by a significant convective activity, shows variations in the radiative variables due to the differences between the original and the NN schemes. The offline results presented in Section \ref{sec:NN} suggest that the NN deep convection scheme smoothes the tendencies of learned deep convective processes (Fig. \ref{fig:zm_offline}). This can account for the underestimation of deep convection processes in the ARP-GEM1-NNDC experiment.  In the ARP-GEM1-noDC simulation, the lack of subgrid scale parameterization has a strong impact on radiative variables, in particular in the LW. The ascending branch of the Hadley-Walker circulation, with increased OLR, indicates a significant reduction in high-level clouds. Overall, these findings reinforce the NN emulator’s potential to replicate the radiative effects of deep convection processes, as evidenced by its close agreement with the refrence experiment result.

\subsection{Runtime comparison}

This paragraph evaluates the numerical cost associated with the implementation of the Fortran/Python interface and the use of a NN to replace part of the deep convection parameterization. We evaluate execution times from one month of simulation from the experiments ARP-GEM1 and ARP-GEM1-NNDC, using DrHook profiling tool \citep{saarinen-2005}. Both simulations used 125 AMD Rome CPU cores to run the GCM. The ARP-GEM1-NNDC experiment also employed 4 Nvidia V100 GPUs for the NN inference, chosen for their faster processing capabilities compared to CPUs, despite the NN’s relatively low complexity.

In this configuration, the simulation ARP-GEM1-NNDC is slightly less than twice as expensive as the reference simulation. As the NN inference runs in parallel with the radiative transfer and surface parameterizations, and since it is less expensive than these parameterizations, it does not contribute direcly to the computational overhead. It is mainly due to the field exchange itself between ARP-GEM1 and the Python component. Future improvements could focus on optimizing the parallelization of the field exchange to reduce this overhead. Note that ARP-GEM1 is computationally highly efficient, hence any additional cost can become expensive relatively to the other components.

\section{Conclusion and discussion}

This study addresses the technical aspects of implementing and evaluating a NN-based parameterization within a GCM, specifically ARP-GEM1. We used the OASIS coupler and its Python interface for field exchanges between ARP-GEM1 and a Python component responsible for executing the NN inference. The key advantage of the 'field-exchange' approach is the ability to run the Python component in parallel with ARP-GEM1, allowing the execution of the GCM and the NN inference on heterogeneous partitions. It also allows the execution of the model using heterogeneous CPU/GPU computational resources. In particular, running the Python component including the NN inference can benefit from substantial speedup when executed on GPUs.  

As a first test case, we trained a simple NN to emulate ARP-GEM1's deep convection parameterization and conducted a 5-year online simulation at 55-km horizontal resolution, with the NN deep convection parameterization. The NN’s online performance was evaluated by comparing key climate variables: TOA radiation fluxes and precipitation. The results of the simulation using the NN deep convection shows good agreement with the reference simulation. 

Whereas the NN inference does not directly increase the cost of the simulation, the computational overhead is due to the field exchange between the GCM and the Python component. Further optimization of the interface would be necessary to reduce the numerical cost of the field exchange. 

The `field-exchange' approach presents significant benefits in integrating NN inference into heterogeneous HPC architectures. By enabling the execution of NN components on GPUs, this method enhances flexibility and delivers considerable speedups, particularly when applied to complex NN parameterizations, such as generative models. The use of GPUs also helps overcome memory limitations associated with larger NN models, which might otherwise hinder their deployment on CPU-only systems, especially when resources are shared with the host GCM.

Furthermore, leveraging GPUs for NN inference can pave the road for more sophisticated applications, such as the development of non-local parameterizations. NNs, especially those utilizing convolutional layers, offer a promising avenue for efficiently integrating non-local information into parameterizations. Due to computational resources limitations, most physical parameterizations are not designed to take into account spatial information.

Although the current implementation of the field-exchange method may introduce some numerical overhead, it allows for the integration of NN inference into the ARP-GEM system with minimal modifications to its Fortran legacy code, while still benefiting from GPU-based acceleration. As the use of GPUs becomes more prevalent in atmospheric modeling, the adoption of fully Python-based models, like NeuralGCM \citep{kochkov_neural_2024}, may become a necessary step forward.

%%%%%%%%%%%%%%%%%%%%%%%%%%%%%%%%%%%%%%%%%%%%%%%%%%%%%%%%%%%%%%%%%%%%%
% ACKNOWLEDGMENTS
%%%%%%%%%%%%%%%%%%%%%%%%%%%%%%%%%%%%%%%%%%%%%%%%%%%%%%%%%%%%%%%%%%%%%
\acknowledgments

%%%%%%%%%%%%%%%%%%%%%%%%%%%%%%%%%%%%%%%%%%%%%%%%%%%%%%%%%%%%%%%%%%%%%
% DATA AVAILABILITY STATEMENT
%%%%%%%%%%%%%%%%%%%%%%%%%%%%%%%%%%%%%%%%%%%%%%%%%%%%%%%%%%%%%%%%%%%%%
% 
%
\datastatement
The supporting dataset and code are available on Zenodo: \url{https://doi.org/10.5281/zenodo.13795198}. The OASIS coupling library is available: \url{https://gitlab.com/cerfacs/oasis3-mct/-/tree/OASIS3-MCT_5.0}.

\bibliographystyle{ametsocV6}

\end{document}